# Gravity surveys using a mobile atom interferometer

Xuejian Wu[1], Zachary Pagel[1], Bola S. Malek[1], Timothy H. Nguyen[1], Fei Zi[1], Daniel S. Scheirer[2], Holger Müller[1,3,*]

[1]Department of Physics, University of California, Berkeley, California 94720, USA
[2]U.S. Geological Survey, 345 Middlefield Road, MS 989, Menlo Park, California 94025, USA
[3]Molecular Biophysics and Integrated Bioimaging, Lawrence Berkeley National Laboratory, Berkeley, California 94720, USA
*Correspondence to: hm@berkeley.edu

## Abstract

Mobile gravimetry is important in metrology, navigation, geodesy, and geophysics. Atomic gravimeters could be among the most accurate mobile gravimeters, but are currently constrained by being complex and fragile. Here, we demonstrate a mobile atomic gravimeter, measuring tidal gravity variations in the laboratory as well as surveying gravity in the field. The tidal gravity measurements achieve a sensitivity of 37 μGal/√Hz (1 μGal=10 nm/s$^2$) and a long-term stability of better than 2 μGal, revealing ocean tidal loading effects and recording several distant earthquakes. We survey gravity in the Berkeley Hills with an accuracy of around 0.04 mGal and determine the density of the subsurface rocks from the vertical gravity gradient. With simplicity and sensitivity, our instrument paves the way for bringing atomic gravimeters to field applications.

## Introduction

Light-pulse atom interferometers (1) have been used to measure inertial forces (2-6) and fundamental constants (7, 8), test fundamental laws of physics (9), and search for physics beyond the standard model (10). Gravimeters based on atom interferometry are among the most accurate tools for measuring gravity (11). By contrast to instruments based on springs (12), superconducting coils (13), microelectromechanical devices (14), or falling corner cubes (15), atomic gravimeters rely on matter-wave interferometry with a freely falling atomic cloud. Matter waves are directed into two interferometer arms by the momentum of photons, extremely well-defined through the laser wavelength. Transportable atomic gravimeters are being developed towards metrology (16-20), airborne sensing (21), shipborne surveys (22), and field applications (23-25). The most sensitive transportable atomic gravimeters reach a sensitivity of 10 to 50 μGal/√Hz (1 μGal=10 nm/s$^2$) in the laboratory (16, 17), but the only atomic gravimeter used in gravity surveys achieves a precision of only ~1 mGal on a ship (22). Meanwhile, precise mobile gravimetry is valuable in broad areas. Gravity measurements with an uncertainty of a few μGal are required for using the Watt balance to realize the new definition of the kilogram (26). The use of gravity reference maps to aid inertial marine navigation requires onboard gravimeters with at least mGal accuracy (27). Seasonal aquifer fluctuations can be monitored by sensing μGal-scale gravity changes (28). These examples illustrate that atomic gravimeters must be not only sensitive, but also mobile and reliable in field conditions.

Here, we demonstrate lab- and field-operation of a mobile atomic gravimeter. We achieve a sensitivity of 37 μGal/√Hz and a stability of better than 2 μGal in half an hour. Comparing the measured gravity with a solid-earth tide model, the atomic gravimeter is sensitive enough to reveal ocean tidal loading effects and to measure seismic waves of distant earthquakes. The atomic gravimeter measures absolute gravity in the laboratory with an uncertainty of 0.02 mGal, confirmed by a spring-based relative gravimeter referencing to a site with known absolute gravity. Furthermore, the mobility allows us to measure gravity in the field with a resolution of around 0.5 mGal/√Hz, depending on environmental noise. We implement gravity surveys in the Berkeley Hills along a route of ~7.6 km and an elevation change of ~400 m. At each static measurement location, it takes about 15 minutes to set up the gravimeter and a few minutes to measure gravity with an uncertainty of around 0.04 mGal. From the measured vertical gravity gradient, the density of subsurface rocks is estimated to be 2.0(2) g/cm$^3$. Geodetic and geophysical studies, such as refining the geoid, resource exploration, hydrological studies, and hazard monitoring, can benefit from precise absolute gravity measurements using field-operating atomic gravimeters.

## Results

### Mobile atomic gravimeter

The mobile atomic gravimeter is based on an atom interferometer, schematically shown in Fig. 1A. It features a magneto-optical trap (MOT) inside a pyramid mirror with a through-hole. This novel geometry offers many advantages. First, it acts as a differential pumping stage between the MOT and atom interferometry regions. A vapor pressure ratio of more than 10:1 (see fig. S1 in Supplementary Materials) accelerates atom-loading speed and decreases background noise in atom detection. We achieve a signal-to-noise ratio of 200:1 (see fig. S1 in Supplementary Materials) and reduce systematic effects from the refractive index of background atoms, particularly important when the laser is at a small detuning (see Materials and Methods). Second, it allows the MOT and interferometer laser beams to have different waists such that we can obtain both a large MOT volume and high Raman-beam intensity with the available laser power. Third, the atomic gravimeter takes advantage of retroreflection from a vibration-isolated mirror, and is insensitive to vibrations of the pyramid mirror. Thus, the vibration isolation is simpler and more effective than in traditional pyramidal atomic gravimeters (23, 24). Finally, using a flat mirror as the retroreflector eliminates the systematic effects from imperfections in the pyramidal top angle and wavefront aberration due to the pyramid edges.

The MOT beam and its reflections off the pyramid mirror and the retroreflector trap ~5×10$^7$ cesium atoms in a 150-ms loading time. Polarization gradient cooling, after switching off the MOT magnetic fields, further cools the atoms to ~2 μK. The atoms are then released to fall freely under gravity. A microwave pulse transfers ~5×10$^6$ atoms from the state $F$=4, $m_F$=0 into $F$=3, $m_F$=0 (where $F$ and $m_F$ are the total spin quantum numbers). A resonant laser pulse clears away atoms left in $F$=4.

Atom interferometry is performed underneath the pyramid mirror using Doppler-sensitive two-photon Raman transitions between the $F$=3 and $F$=4 hyperfine ground states, driven by two laser beams with wave vectors $\mathbf{k}_1$ and $\mathbf{k}_2$. We use a Mach-Zehnder geometry, as shown in Fig. 1B. A $\pi/2$ pulse can place the atoms into a superposition of the two states. A $\pi$ pulse can fully transfer atoms from one state to the other. The three pulses are equally spaced by a pulse separation time $T$ and their durations are 4, 8, and 4 μs, respectively. Since the atoms move in free-fall, they see a Doppler-shifted laser frequency. To compensate this effect, we linearly ramp the laser frequency difference between the two beams with a rate of $\alpha$.

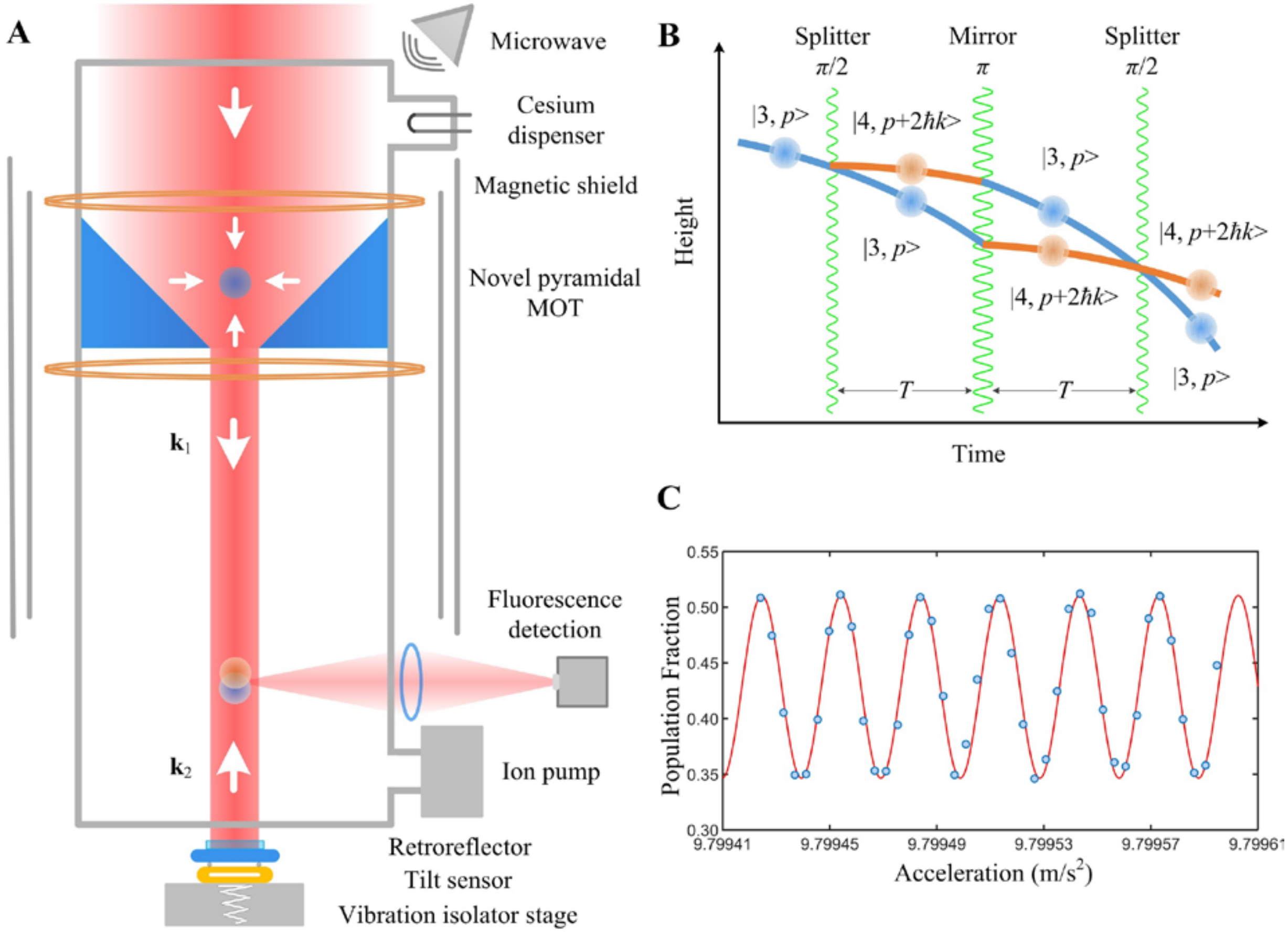

**Fig. 1. Atomic gravimeter.** (**A**) Schematic. Cesium clouds are loaded in the novel pyramidal MOT and then freely fall into the region of fluorescence detection. $\mathbf{k}_1$ and $\mathbf{k}_2$ are the wave vectors of the interferometer beams. A magnetic shield and a solenoid (not shown) around the vacuum chamber create a uniform magnetic bias field. The retroreflector consists of a flat mirror and a quarter-wave plate. The vibration isolation stage includes a passive vibration isolation table, a seismometer, voice coils and an active feedback loop. (**B**) Mach-Zehnder interferometer geometry. Three laser pulses (wavy green lines) split, redirect, and combine a matter wave (blue and orange lines). (**C**) Fringes with $T$=120 ms and $C$=16%. The blue dots are single-shot experimental data and the red curve is a sinusoidal fit.

The population fraction of the atoms in the two states can be used to estimate the phase difference $\Delta\varphi$ between the two interferometer arms, $P=P_0+(C/2)\cos(\Delta\varphi)$, where $P_0$ is the normalized background population and $C$ is the contrast. For atomic gravimeters, the phase difference $\Delta\varphi=(\mathbf{k}_{\mathrm{eff}}\cdot\mathbf{g}-\alpha)T^2$, where $\mathbf{k}_{\mathrm{eff}}=\mathbf{k}_1-\mathbf{k}_2$ is the effective wave vector of the interferometer beams and $\mathbf{g}$ is the gravitational acceleration. By varying $\alpha$, the interference fringes, and thus the acceleration, can be obtained (Fig. 1C).

The $\cos(\Delta\varphi)$-dependence of the fringes introduces an ambiguity in the gravity measurement that we resolve by using different pulse separation time. To measure absolute gravity, we start with a short $T$ of several ms to obtain a unique but coarse measurement, and process finer measurements with longer $T$. For each $T$, we reverse the wave vectors to reduce the systematic effects from the magnetic gradient and the first-order light shift (29). This procedure is performed automatically at the start of a new measurement (see fig. S2 in Supplementary Materials).

All laser beams necessary for the MOT, interferometry, and detection are generated from a single diode laser using three acousto-optic modulators and one fiber-based electro-optic modulator (see Materials and Methods and fig. S3 in Supplementary Materials). The 50-mW cooling beam has a waist of 13 mm ($1/e^2$ radius) and the 25-mW Raman beam has a waist of 5 mm. They are combined before entering the vacuum chamber. A liquid crystal retarder switches between circular and linear polarization. For trapping atoms, counterpropagating $\sigma^+/\sigma^-$ polarization pairs are formed inside the pyramid thanks to the pyramidal geometry. For atom interferometry, lin $\perp$ lin polarization is used. The single-photon detuning of the Doppler-sensitive two-photon Raman transition is -158 MHz relative to $F$=4→$F'$=5. At this detuning, we obtain a $\pi$ pulse of as short as 8 μs. Thanks to the fast Rabi frequency, more than 60% of the total atoms can be addressed without velocity selection of the atomic cloud. With an effect of all the Raman laser sidebands, this detuning can also cancel the differential AC Stark shift, strongly reducing systematic effects (24). For $T$=10 ms, we obtain a fringe contrast higher than 30%. For $T$=120 ms, we achieve fringe contrast of 16% (Fig. 1C), which may be limited by the broad velocity width of the atomic cloud and the inhomogeneous Rabi flopping of the interferometer pulses across the cloud.

**Tidal gravity measurements**

Figure 2 illustrates the long-term tidal gravity variation measured over 12 days by the atomic gravimeter. We operated the atom interferometer with $T$=130 ms and active vibration isolation (see Materials and Methods and fig. S4 in Supplementary Materials). The cycle time was 0.481 s. Fitting an interference fringe with 16 drops, one gravity measurement took 7.696 s (see fig. S5 in Supplementary Materials for the gravity data per fringe). Figure 2A shows the data averaged over every 2 hours compared with a solid-earth tide model (30).

Because our laboratory is about 4.5 km east of the San Francisco Bay, the ocean tidal loading effect on gravity at our location is significant and is not accurately described in available models (31). Figure 2B shows that the residual gravity variation is correlated to the water-level variation measured in the Bay. As shown in Fig. 2C and 2D, the ocean tidal loading leads to the peaks in the Allan deviation and the power spectral density in a period-band between 6-24 hours. After correcting for the solid-earth tide, the sensitivity of the atomic gravimeter is 37 μGal/√Hz and the stability in half an hour is better than 2 μGal (Fig. 2C).

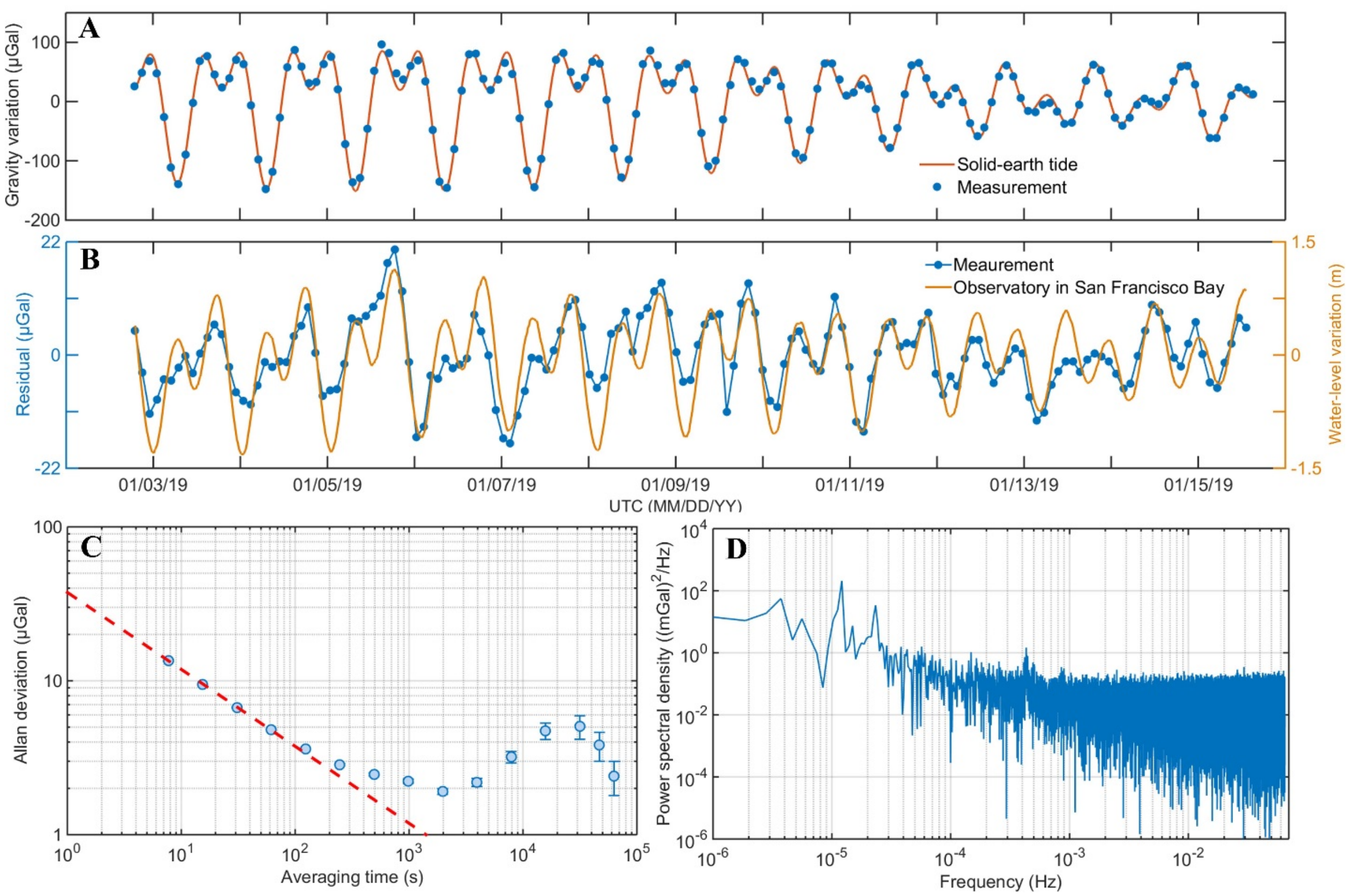

**Fig. 2. Tidal gravity measurement.** (**A**) Tidal gravity variation as a function of time. (**B**) Comparison between the gravity residual and the water-level variation in the San Francisco Bay. The gravity residual is the difference between the measurements and the solid-earth tide model. The water-level variation is measured by the observatory of National Oceanic and Atmospheric Administration in Richmond, California. (C) Allan deviation of the residual. The dashed line corresponds to a sensitivity of 37 μGal/√Hz. The broad peak around $3\times10^4$ s is due to the ocean tidal loading. (D) Power spectral density of the residual. The ocean tidal loading results in the peaks around $1\text{-}3\times10^{-5}$ Hz.

During the tidal gravity measurements, the atomic gravimeter recorded seismic wave trains from several distant earthquakes. Because the frequency of the seismic waves generated by the earthquakes was much lower than the resonant frequency of the active vibration isolation, the ground motion passed through linearly and moved the retroreflector vertically. Thus, the atomic gravimeter measured the vertical acceleration of the seismic waves. Figure 3 illustrates a comparison between the atomic gravimeter and one of seismometers in the Berkeley Digital Seismic Network. On January 5 2019, a 6.8-magnitude and 570-km-deep earthquake occurred in Brazil at 19:25 UTC. After about 20 minutes, waves from this earthquake detected by both the atomic gravimeter and the seismometer. On January 6 2019, a 6.6-magnitude and 43-km-deep earthquake occurred in Indonesia at 17:27 UTC, and its dispersive Rayleigh wave arrived at Berkeley around 18:16 UTC. According to the measured acceleration as a function of time, the vertical component of the Rayleigh wave had a period of ~30 s and a peak-to-peak amplitude of ~90 μm.

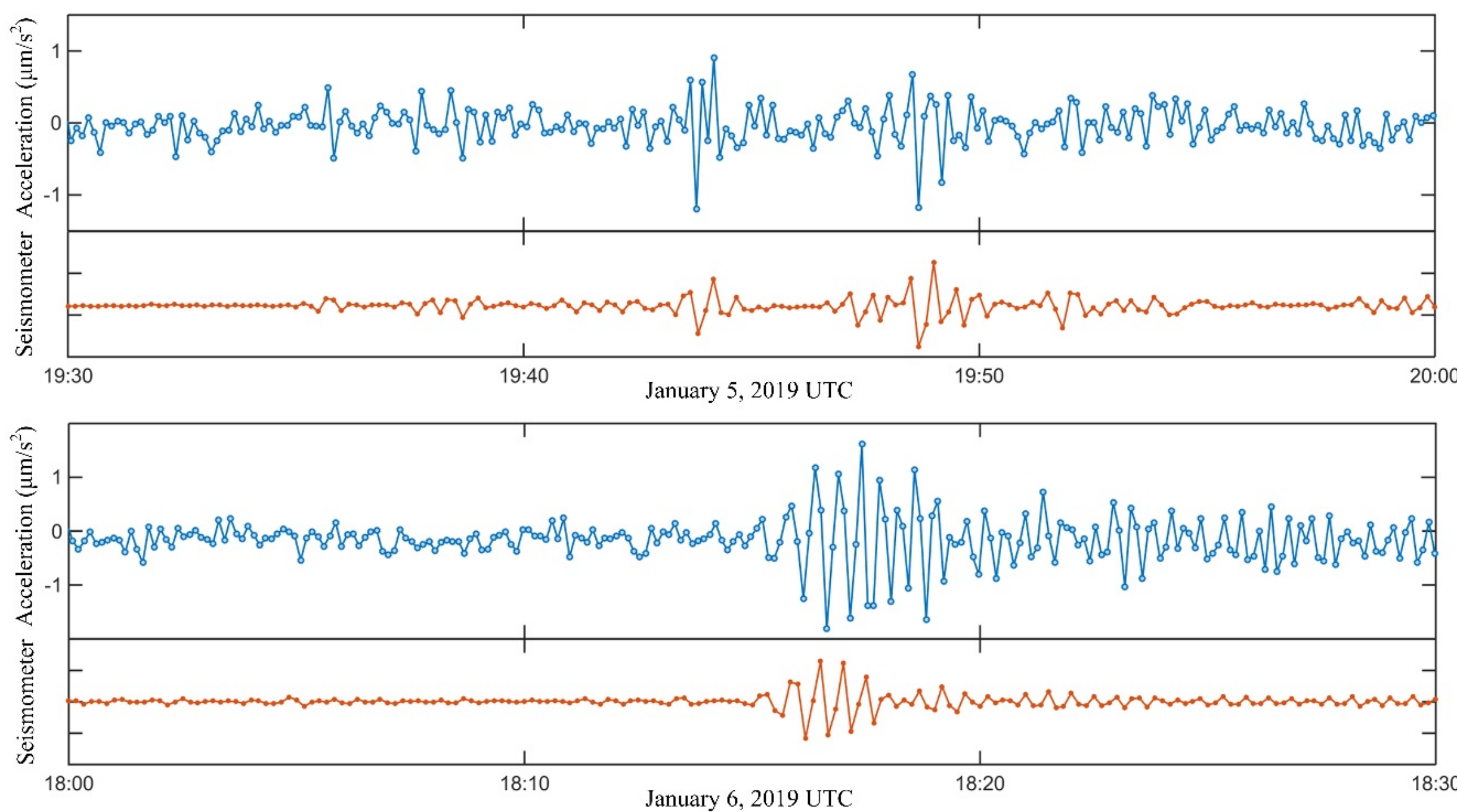

**Fig. 3. Earthquake seismic waves detected in Berkeley.** The atomic gravimeter measures the vertical acceleration of the seismic waves with an update rate of 0.13 Hz. The seismic signal is the vertical channel of the seismometer located in Haviland Hall on the UC Berkeley campus. It is in arbitrary unit and has an update rate is 0.1 Hz.

**Systematic effects and repeatability**

To investigate the accuracy of the atomic gravimeter, we estimated the systematic effects. With major errors from the magnetic fields and the refractive index of background vapor, the total systematic error is 0.015 mGal and the measurement bias is -0.010 mGal (see Materials and Methods and table S1 in Supplementary Materials for details).

To verify the repeatability after transporting the atomic gravimeter, we measured gravity on different floors of Campbell Hall on the UC Berkeley campus, comparing with the gravity differences measured by a spring-based relative gravimeter (CG-5, Scintrex). The atomic gravimeter measured 979 955.61(2) mGal on the floor of the basement (in the laboratory). This value matches one estimated by the relative gravimeter, 979 955.58(1) mGal, using standard gravity surveying techniques (32) and referencing to a gravity station with known absolute gravity, on the USGS campus in Menlo Park, California.

We took fringes with $T$=70 ms for the measurements on the upper floors. It had an unambiguous range of 8.7 mGal, larger than the gravity variation in the building. Because the seismometer in the active vibration isolation needs several hours to warm up after transport, we did not use the active system but only the passive stage. The atomic gravimeter achieved a sensitivity of around 0.2 mGal/√Hz depending on vibrational noise. Though the sensitivity on higher floors decreased due to stronger vibration noise, the gravimeter took data for several minutes on each floor and thus ensured a statistical uncertainty below 0.05 mGal. The experiment of measuring gravity on the floors of the building was carried out by only one person within 3 hours, most of the time spent on transporting and realigning to the vertical axis (see Materials and Methods).

Taking the basement-floor gravity as a reference, Fig. 4 compares the gravity variation from the atomic gravimeter and the relative gravimeter (see table S2 in Supplementary Materials for the absolute gravity on each floor). We excluded the gravity measured on the basement floor when fitting the gravity gradient because the gradient is expected to be significantly different below and above the ground level. The atomic gravimeter measures a vertical gravity gradient of -0.289(3) mGal/m, and the relative gravimeter obtains -0.285(1) mGal/m, matching within the statistical error. These gradients are smaller than the free-air gradient (-0.3086 mGal/m), indicating the gravitational effect of the mass of the building.

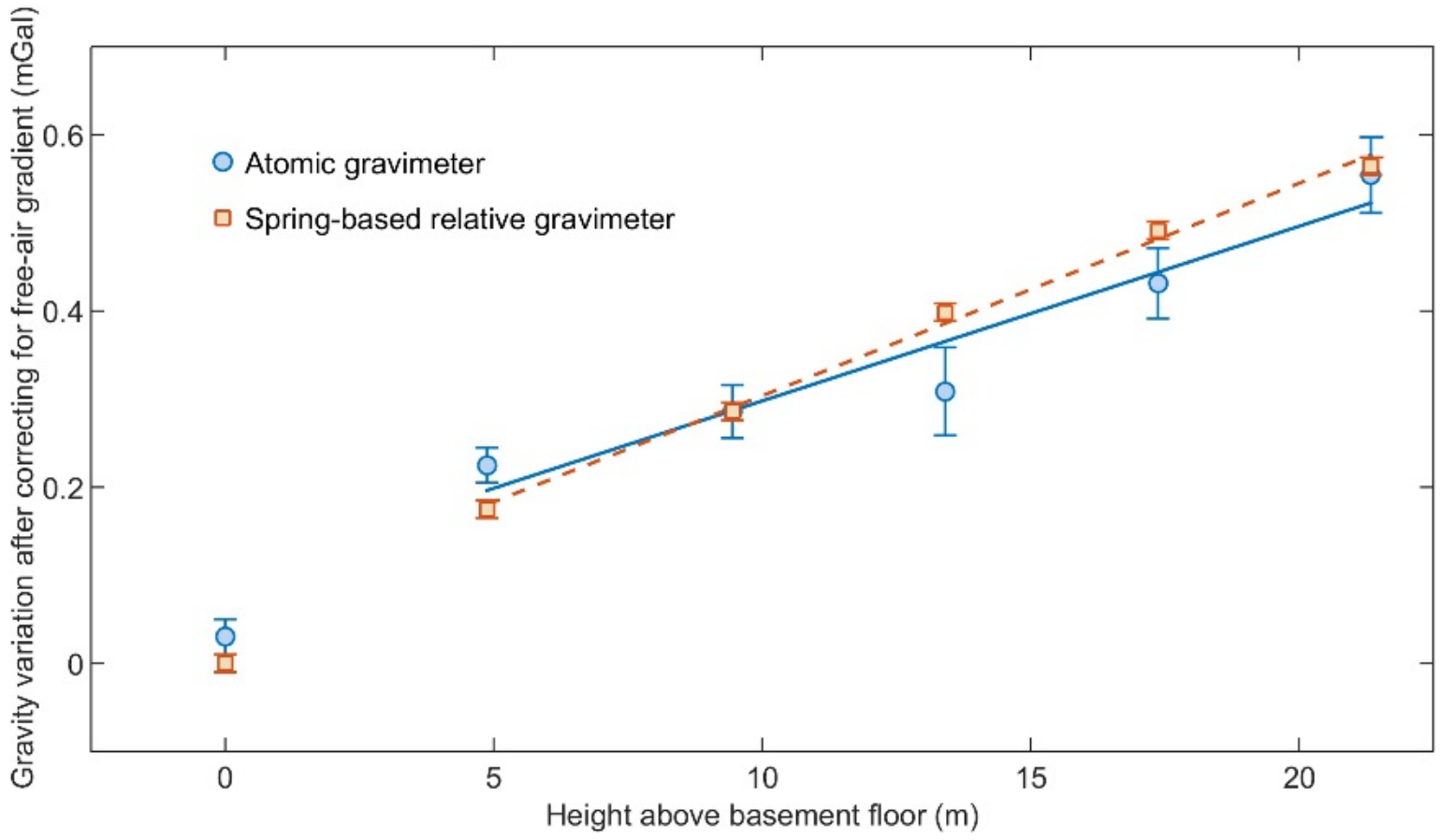

**Fig. 4. Gravity variation on different floors of Campbell Hall.** The data at 0 m is the gravity measured on the floor of the basement and the others are from floors 1 to 5. The error bars are 1-$\sigma$ statistical and systematic errors. The height of the floors is obtained from the building design. The free-air gradient of -0.3086 mGal/m is removed from the data to emphasize deviations of the vertical gravity gradient from the free-air value. The solid line is a linear fit of the gravity variation measured by the atomic gravimeter from floors 1 to 5, and determines a vertical gravity gradient of -0.289(3) mGal/m. Similarly, the dashed line is a linear fit of the relative gravimeter, determining a vertical gravity gradient of -0.285(1) mGal/m.

**Gravity survey in Berkeley Hills**

To demonstrate the use of the atomic gravimeter in the field, we measured absolute gravity in the Berkeley Hills. As shown in Fig. 5A, the route had a length of ~7.6 km and an elevation change of ~400 m. We operated the atomic gravimeter inside a vehicle using passive vibration isolation and measured gravity at 6 locations. At each location, it took about 15 minutes to set up the gravimeter, including powering-up the instrument and aligning the interferometer beam to the gravity axis (see Materials and Methods). Due to increased vibrational noise in the field, the sensitivity of the gravimeter was around 0.5 mGal/√Hz. We first determined the absolute gravity with $T$=10 ms and increased with steps of 10 ms, and finally took more fringes with $T$=70 ms to reach a statistical uncertainty of around 0.03 mGal. A microcontroller system managed this procedure automatically (see Materials and Methods). Considering the systematic effects, the uncertainty of the gravity values was around 0.04 mGal. The laser system required no realignment during the 7-hour survey, with air temperature changing by 12 °C and significant transport vibrations.

The gravity changes by -92.6 mGal from the base to the peak of the Berkeley Hills (see table S3 in Supplementary Materials for the absolute gravity at each measurement location). To determine the vertical gravity gradient, the measured gravity is corrected for solid-earth tides, latitude variations, and terrain effects (see Materials and Methods). Figure 5B illustrates the gravity anomaly as a function of the elevation. The vertical gravity gradient is -0.225(10) mGal/m, a value that is 0.084(10) mGal/m smaller than the free-air gradient because of the mass of the rocks forming the Berkeley Hills. This gradient yields an average density value of the Berkeley Hills of 2.0(2) g/cm$^3$. The vertical gravity gradient and the average density match those calculated from USGS gravity stations collected in 1998 along the same path.

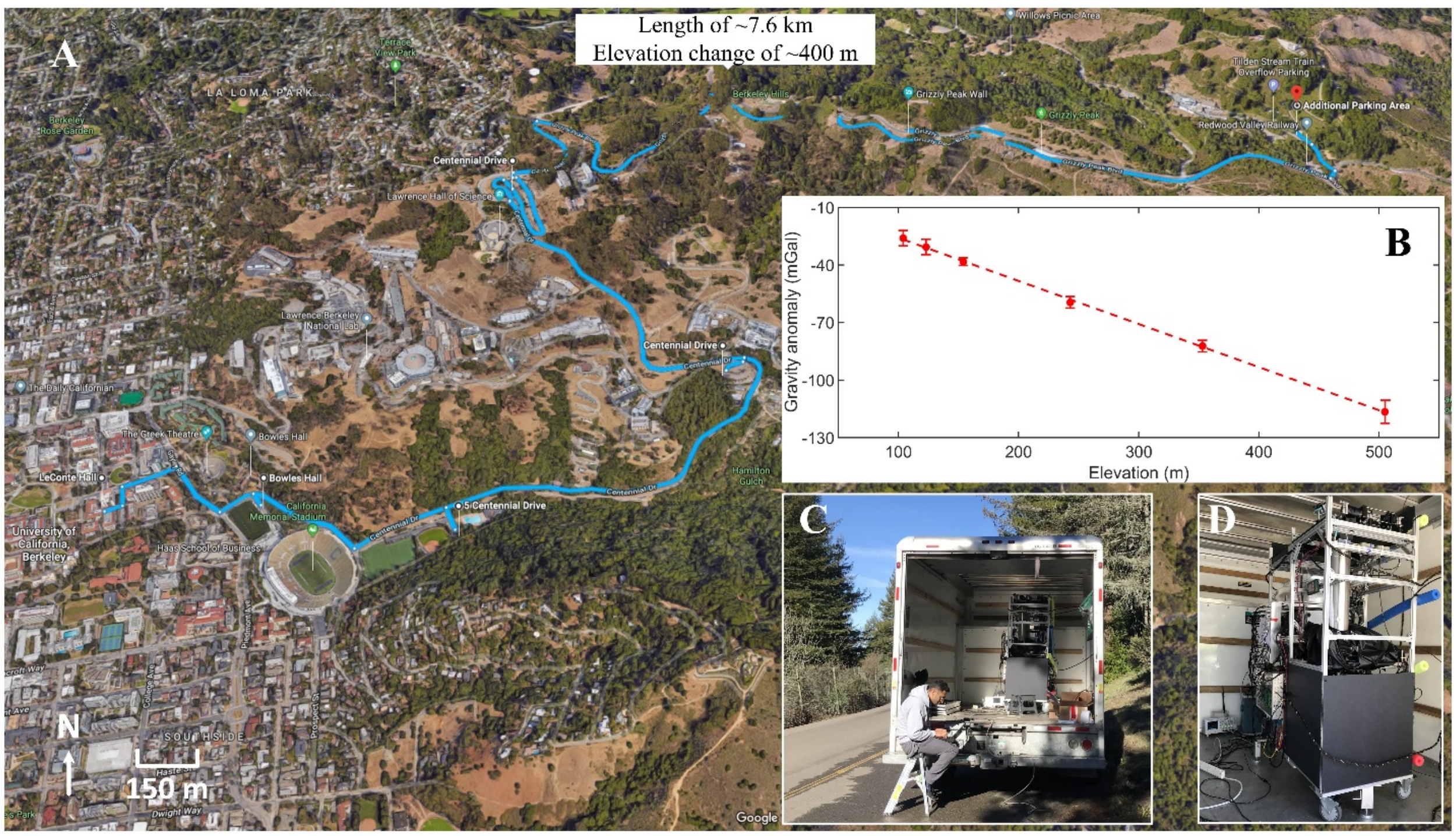

**Fig. 5. Gravity survey in Berkeley Hills.** (**A**) Measurement route. The blue curve depicts the route and the white pin drops are the 6 measurement locations. (**B**) Gravity anomaly as a function of the elevation. Elevations are from Google maps. The error bars are 1-$\sigma$ statistical and systematic errors. The dashed line indicates a vertical gravity gradient of -0.225(10) mGal/m. (**C**) Field-operation of the atomic gravimeter inside a vehicle. (**D**) The atomic gravimeter apparatus.

## Discussion

We have developed a mobile atomic gravimeter and performed tidal gravity measurements and gravity surveys. Our instrument uses a novel pyramidal MOT that takes advantage of single-beam atom interferometry and offers differential pumping, simple laser-to-gravity alignment, and enhanced vibration isolation. Our atomic gravimeter is mobile, compact, and robust over transport in the field while maintaining comparable sensitivity to other state-of-the-art transportable atomic gravimeters (16-25). These features make it a candidate for geodetic and geophysical applications requiring precise mobile gravimetry (33).

The sensitivity of the atomic gravimeter is currently limited by vibrational noise. However, the sensitivity as a function of the pulse separation time shows the tendency that a longer pulse separation time would still help (see fig. S4 in Supplementary Materials). Hence, we can further improve the sensitivity by dropping the atomic cloud longer. Because the local gravity is affected by the tidal effects, the inaccurate tide model at our location constrains the accuracy of the long-term stability measurement and the systematic effect evaluation. A gravity comparison at a geophysical observatory would allow us to characterize them more accurately (34). With these improvements, a more accurate measurement of the ocean tidal loading effect may be useful for investigating the Earth's mass structure and its variation with time at levels beyond current precision (31, 33). In addition, atomic gravimeters with mobility, sensitivity, and accuracy may find more applications in detecting tunnels, sensing underground water storage, and monitoring earthquake and volcano activity.

## Materials and Methods

### Apparatus

The atomic gravimeter is installed in a 1 m × 0.8 m × 1.7 m (L×W×H) cart (Fig. 5D). It weighs around 100 kg, mostly due to the lithium battery power supply, vibration isolation stage, and cart. The cart has two columns, one for the electronic system and the other for the vacuum system. The laser unit consists of two 60 cm × 46 cm optical breadboards placed at the top level of the cart. The total power consumption is about 250 W. For field operation, the gravimeter is powered by a 1450-Wh lithium power station (Yeti 1400, Goal Zero). A gas generator is used as a backup power supply. The vehicle used for the gravity surveys is a 15-ft truck (Fig. 5C).

### Pyramid mirror

The pyramid mirror is in a glass cube of 40 mm × 40 mm × 40 mm (Fig. 1A). The diameter of the center hole is 10 mm. Its inner faces are coated with protected gold in order to generate equal phase shift between two orthogonal polarizations for the reflections. The pyramid mirror is attached to an aluminum holder and fixed to the bottom vacuum flange by rods. Its outside dimension is designed to fit within the vacuum chamber with a tolerance of ±0.5 mm.

### Vacuum chamber

The vacuum chamber is a 0.6-m-long glass cylinder with an outside diameter of ~60 mm (fig. S1). There is one 0.1-m-long rectangular cell near each end. They are separated by 0.3 m. The pyramid mirror is in the top cell and the atoms are detected in the bottom cell. The pressure in the interferometry region is $1\times10^{-9}$ torr measured by the ion pump (5S, Gamma Vacuum), of which ~$3\times10^{-10}$ torr is cesium vapor (when the dispenser is off, the residual pressure is $7\times10^{-10}$ torr). The cesium pressure in the MOT region is ~$1\times10^{-8}$ torr estimated by fluorescence detection.

## Laser system

The laser system (fig. S3) uses only one 240-mW distributed Bragg reflector diode laser (PH852DBR, Photodigm) with 852-nm wavelength and 500-kHz linewidth. The laser is frequency-stabilized by polarization spectroscopy in a cesium vapor cell, nominally to the $F=4\rightarrow F'=4/F=4\rightarrow F'=5$ crossover line. An acousto-optical modulator (AOM) can vary the laser frequency relative to the spectroscopy by ±15 MHz to switch between frequencies needed for trapping the atoms and for detecting them after interferometry. A 125-MHz frequency change is available by adding a bias voltage to the spectroscopy signal, which changes the lock point to the $F=4\rightarrow F'=4$ line. This is used to generate a ~140-MHz red detuning for polarization gradient cooling. Another AOM acts as an overall on/off switch for the beam and shifts it to about -10 MHz from $F=4\rightarrow F'=5$ to act as the cooling beam. The beam is coupled into a polarization maintaining fiber. A fraction of the beam is deflected by another AOM and sent through a fiber-coupled electro-optical phase modulator (EOM), which generates the repumping frequency as a sideband. Though the repumping beam has a smaller waist than the cooling beam, we do not observe a decrease in the number of atoms. For interferometry pulses, most power is directed through the EOM, and the carrier and one sideband serve as the frequency pair to drive Raman transitions. The high peak power does not degrade the EOM lifetime because of the short pulse time. Fluorescence detection is performed using the MOT laser.

## Electronic system

The electronics system includes a microcontroller system, drivers for the laser and modulators, and a microwave frequency chain. The microwave frequency chain is a phase-lock loop that stabilizes the microwave frequency of a dielectric resonant oscillator to the high-order harmonic of a phase-locked radio-frequency oscillator. By varying the offset frequency, the microwave frequency chain can produce the frequency of the cesium clock transition as well as ramp the frequency to compensate for the Doppler shift of the freely-falling atomic clouds. The electronic system refers to a 10-MHz Rubidium clock (PRS10, Stanford Research System). The microcontroller system generates all the controlling digital and analog signals, and measures the analog signal from the photodetector to obtain an interference fringe. A detailed implementation of the microcontroller system can be found in Ref. (35).

## Vibration isolation

The vibration isolation stage includes a passive vibration isolation table (25BM-10, Minusk), a broadband seismometer (MBB-2, Metrozet), two voice coils inside the table and a digital feedback loop. The digital feedback loop is a finite impulse response (FIR) filter based on a field-programmable gate array (FPGA). The output from the seismometer is amplified by a factor of 200, and then sampled into the FPGA (STEMlab 125-14, Red Pitaya) with a rate of ~3.8 kS/s and a resolution of 14 bit. Based on Ref. (36), the FIR filter is designed to suppress vibrations around the resonance frequency of the Minusk stage, roughly around 1 Hz. Around this frequency, the transfer function of the digital filter includes elements to enhance the gain of the feedback loop. Additional filter elements are added to provide stability near the 0.01-Hz unity gain point as well as mechanical resonances above 100 Hz. An in-loop measurement shows a 500-fold suppression in vibration noise around 1 Hz (fig. S4).

**Alignment procedure**
Before measuring gravity, we first adjust the leveling feet of the cart using a bubble level on top of the vacuum chamber. Second, we adjust the leveling feet under the vibration isolation table using a bubble level on the seismometer. Third, we adjust the retroreflector using an electronic tilt sensor (700-series, Applied Geomechanics) with a resolution of 1 µrad. Fourth, we align the interferometer beam to be perpendicular to the retroreflector by adjusting the pair of mirrors after the telescope. We achieve this by overlapping the reflection beam and the incident beam through a pinhole, with an accuracy of about 50 µrad. Last, we adjust the cooling beam to center the atomic cloud with respect to the pyramid hole.

**Systematic effects**
The total systematic error of our mobile atomic gravimeter is 15 µGal and the measurement bias is -10.3 µGal (table S1). Here, we briefly discuss those contributions that produce an error larger than 5 µGal.

Magnetic fields. The cesium ground state hyperfine splitting exhibits a quadratic Zeeman effect of 0.43 kHz/G$^2$. A homogenous magnetic field has no influence on the phase of a Mach-Zehnder atom interferometer, but a magnetic-field gradient $B'$ on top of a bias field $B_0$ can add an effective force that is proportional to $B_0B'$. To characterize the shift, we vary $B_0$ and find that the gravity measurement is modified by ~1.8 mGal/G. The wavevector-reversal reduces it to (-9±28) µGal/G (fig. S6). At a nominal bias field of 0.3 G, this systematic is (-3±9) µGal.

Refractive index of background vapor. Operating with small detunings, we have to check whether the refractive index of cesium background atoms generates a substantial systematic error. It is related to the atomic polarizability $\alpha$, and thus to the AC Stark shift $\omega^{AC}$, by $(n\text{-}1)=2\pi\alpha\rho$ $=2\pi c\rho\hbar\omega^{AC}/I$, where $\rho$ is the number density of atoms, $c$ is the speed of light in vacuum, and $I$ is the laser intensity. The calculation of the AC Stark shift can be found in our previous paper (24). This refractive index is averaged over the $F$=3 and $F$=4 ground states (background atoms are equally distributed in either state), the two Raman frequencies, and the 270-MHz-wide thermal distribution of the Doppler shifts. With a cesium partial pressure of $(3\pm3)\times10^{-10}$ torr, the index of refraction is $(n\text{-}1)=(-7.3\pm7.3)\times10^{-9}$, corresponding to an error in the gravity measurement of (-7.3±7.3) µGal.

Coriolis effect. The Coriolis effect causes an extra phase shift of $2\mathbf{\Omega}\cdot(\mathbf{v}\times\mathbf{k}_{\mathrm{eff}})T^2$ and thus an error $\Delta g/g=2\mathbf{\Omega}\cdot(\mathbf{v}\times\mathbf{k}_{\mathrm{eff}})/(\mathbf{k}_{\mathrm{eff}}\cdot\mathbf{g})$, where **v** is the initial velocity of the atoms and $\mathbf{\Omega}$ is the Earth's rotation. The Earth's rotation at the latitude of Berkeley has a horizontal component of 58 µrad/s. For a transverse velocity component of the atoms of 0±1 mm/s, the systematic is ±6 µGal.

Vertical alignment. To calibrate the alignment to the gravity axis, we measure gravity with different tilt angles of the retroreflector and fit the data with a bivariate quadratic function to obtain the gravity correction (fig. S7). The standard error of the fit residual is ±5 µGal. For each measurement, the gravity values are corrected based on the real-time tilt and the fit function. After correction, the systematic is ±5 µGal.

Aperture effects. The intensity variation of the interferometer beam caused by aperture effects of the pyramidal hole is about 5% with a correlation length in the range of 100 to 200 μm, measured by a beam profiler. This effect on the accuracy of atomic gravimeters has been analyzed for 15% intensity variations and a similar correlation length (37). It shows that such effects decreases rapidly to below $10^{-9}$ as the thermal atom velocity exceeds a few mm/s, because atoms average over high- and low-intensity spots. The thermal expansion speed of our atomic cloud is about 1 cm/s, so that the potential position dependent phase shifts caused by the intensity ripple are negligible.

**Latitude and terrain correction**

We correct the gravity values collected in the Berkeley Hills for latitude variations using the WGS84 ellipsoidal gravity formula (38) to create latitude-corrected gravity anomalies. Further, we correct these gravity anomalies for the effects of terrain using a National Elevation Database 1-arc second (~30 m) digital elevation model and the Plouff approach (39). The terrain correction is calculated in an annulus around a gravity station, with an inner radius of 50 m and an outer radius of 166.7 km, per standard practice. Most of the variability in the terrain correction among these stations arises from the terrain variability within 5-10 km of the stations. The local terrain is rougher at higher elevations in the Berkeley Hills than near their base, and local terrain effects always add to gravity (32). Hence, accounting for the terrain will yield a vertical gravity gradient (VGG) that is smaller than the measured gradient and that can be used to estimate the density of the terrain via the Nettleton method (40). We first calculate terrain corrections using the standard density reduction value of 2.67 $g/cm^3$ to modify the measured VGG. This terrain-corrected VGG implies a density of the Berkeley Hills to be about 2.0(2) $g/cm^3$. Then, we re-calculate the terrain corrections using that density to get a new terrain-corrected VGG that again implies a density value of 2.0(2) $g/cm^3$.

**References and notes**

1. G. M. Tino, M. A. Kasevich, Eds., Atom Interferometry (Proceedings of the International School of Physics “Enrico Fermi,” Course CLXXXVIII, Societa Italiana di Fisica and IOS Press, 2014).
2. A. V. Rakholia, H. J. McGuinness, G. W. Biedermann, Dual-axis high-data-rate atom interferometer via cold ensemble exchange. Phys. Rev. Applied 2, 054012 (2014).
3. K. S. Hardman, P. J. Everitt, G. D. McDonald, P. Manju, P. B. Wigley, M. A. Sooriyabandara, C. C. N. Kuhn, J. E. Debs, J. D. Close, N. P. Robins, Simultaneous precision gravimetry and magnetic gradiometry with a Bose-Einstein condensate: a high precision, quantum sensor. Phys. Rev. Lett. 117, 138501 (2016).
4. G. W. Hoth, B. Pelle, S. Riedl, J. Kitching, E. A. Donley, Point source atom interferometry with a cloud of finite size. Appl. Phys. Lett. 109, 071113 (2016).
5. M. Xin, W. S. Leong, Z. Chen, S.-Y. Lan, An atom interferometer inside a hollow-core photonic crystal fiber. Sci. Adv. 4, e1701723 (2018).

6. D. Savoie, M. Altorio, B. Fang, L. A. Sidorenkov, R. Geiger, A. Landragin, Interleaved atom interferometry for high-sensitivity inertial measurements. Sci. Adv. 4, eaau7948 (2018).
7. G. Rosi, F. Sorrentino, L. Cacciapuoti, M. Prevedelli, G. M. Tino, Precision measurement of the Newtonian gravitational constant using cold atoms. Nature 510, 518-521 (2014).
8. R. H. Parker, C. Yu, W. Zhong, B. Estey, H. Müller, Measurement of the fine-structure constant as a test of the standard model. Science 360, 191-195 (2018).
9. T. Kovachy, P. Asenbaum, C. Overstreet, C. A. Donnelly, S. M. Dickerson, A. Sugarbaker, J. M. Hogan, M. A. Kasevich, Quantum superposition at the half-metre scale. Nature 528, 530-533 (2015).
10. M. Jaffe, P. Haslinger, V. Xu, P. Hamilton, A. Upadhye, B. Elder, J. Khoury, H. Müller, Testing sub-gravitational forces on atoms from a miniature in-vacuum source mass. Nat. Phys. 13, 938-942 (2017).
11. Z.-K. Hu, B.-L. Sun, X.-C. Duan, M.-K. Zhou, L.-L. Chen, S. Zhan, Q.-Z. Zhang, J. Luo, Demonstration of an ultrahigh-sensitivity atom-interferometry absolute gravimeter. Phys. Rev. A 88, 043610 (2013).
12. M. Lederer, Accuracy of the relative gravity measurement. Acta Geodyn. Geomater. 6, 383–390 (2009).
13. J. M. Goodkind, The superconducting gravimeter. Rev. Sci. Instrum. 70, 4131-4152 (1999).
14. R. P. Middlemiss, A. Samarelli, D. J. Paul, J. Hough, S. Rowan, G. D. Hammond, Measurement of the Earth tides with a MEMS gravimeter. Nature 531, 614-617 (2016).
15. T. M. Niebauer, G. S. Sasagawa, J. E. Faller, R. Hilt, F. Klopping, A new generation of absolute gravimeters. Metrologia 32, 159-180 (1995).
16. C. Freier, M. Hauth, V. Schkolnik, B. Leykauf, M. Schilling, H. Wziontek, H.-G. Scherneck, J. Müller, A. Peters, Mobile quantum gravity sensor with unprecedented stability. J. Phys. Conf. Ser. 723, 012050 (2016).
17. B. Fang, I. Dutta, P. Gillot, D. Savoie, J. Lautier, B. Cheng, C. L. Garrido Alzar, R. Geiger, S. Merlet, F. Pereira Dos Santos, A. Landragin, Metrology with atom interferometry: inertial sensors from laboratory to field applications. J. Phys. Conf. Ser. 723, 012049 (2016).
18. S.-K. Wang, Y. Zhao, W. Zhuang, T.-C. Li, S.-Q. Wu, J.-Y. Feng, C.-J. Li, Shift evaluation of the atomic gravimeter NIM-AGRb-1 and its comparison with FG5X. Metrologia 55, 360-365 (2018).
19. X. Zhang, J. Zhong, B. Tang, X. Chen, L. Zhou, P. Huang, J. Wang, M. Zhan, Compact portable laser system for mobile cold atomic gravimeters. Appl. Opt. 57, 6545-6551 (2018).
20. Z. Fu, Q. Wang, Z. Wang, B. Wu, B. Cheng. Q. Lin, Participation in the absolute gravity comparison with a compact cold atom gravimeter. Chin. Opt. Lett. 17, 011204 (2019).
21. R. Geiger, V. Ménoret, G. Stern, N. Zahzam, P. Cheinet, B. Battelier, A. Villing, F. Moron, M. Lours, Y. Bidel, A. Bresson, A. Landragin, P. Bouyer, Detecting inertial effects with airborne matter-wave interferometry. Nat. Commun. 2, 474 (2011).

22. Y. Bidel, N. Zahzam, C. Blanchard, A. Bonnin, M. Cadoret, A. Bresson, D. Rouxel, M. F. Lequentrec-Lalancette, Absolute marine gravimetry with matter-wave interferometry. Nat. Commun. 9, 627 (2018).
23. V. Ménoret, P. Vermeulen, N. Le Moigne, S. Bonvalot, P. Bouyer, A. Landragin, B. Desruelle, Gravity measurements below $10^{-9}$g with a transportable absolute quantum gravimeter. Sci. Rep. 8, 12300 (2018).
24. X. Wu, F. Zi, J. Dudley, R. J. Bilotta, P. Canoza, H. Müller, Multiaxis atom interferometry with a single-diode laser and a pyramidal magneto-optical trap. Optica 12, 1545-1551 (2017).
25. F. Zi, X. Zhang, M. Huang, N. Li, K. Huang, X. Lu, A compact atom interferometer for field gravity measurements. Laser Phys. 29, 035504 (2019).
26. J. O. Liard, C. A. Sanchez, B. M. Wood, A. D. Inglis, R. J. Silliker, Gravimetry for Watt balance measurements. Metrologia 51, S32-S41 (2014).
27. H. Wang, L. Wu, H. Chai, L. Bao, Y. Wang, Location accuracy of INS/gravity-integrated navigation system on the basis of ocean experiment and simulation. Sensors 17, 2961 (2017).
28. K.-H. Chen, C. Hwang, L.-C. Chang, C.-C. Ke, Short-time geodetic determination of aquifer storage coefficient in Taiwan. J. Geophys. Res. 123, 10987-11015 (2018).
29. A. Gauguet,T. E. Mehlstäubler, T. Lévèque, J. Le Gouët, W. Chaibi, B. Canuel, A. Clairon, F. Pereira Dos Santos, A. Landragin, Off-resonant Raman transition impact in an atom interferometer. Phys. Rev. A 78, 043615 (2008).
30. M. van Camp, P. Vauterin, Tsoft: graphical and interactive software for the analysis of time series and Earth tides. Comput. Geosci. 31, 631-640 (2005).
31. N. K. Pavlis, S. A. Holmes, S. C. Kenyon, J. K. Factor, The development and evaluation of the Earth gravitational model 2008 (EGM2008). J. Geophys. Res. 117, B04406 (2012).
32. R. J. Blakely, Potential theory in gravity and magnetic applications (Cambridge Univ. Press, 1996).
33. M. van Camp, O. de Viron, A. Qatlet, B. Meurers, O. Francis, C. Caudron, Geophysics from terrestrial time-variable gravity measurements. Rev. Geophys. 55, 938-992 (2017).
34. D. Schmerge, O. Francis, J. Henton, D. Ingles, D. Jones, J. Kennedy, K. Krauterbluth, J. Liard, D. Newell, R. Sands, A. Schiel, J. Silliker, D. van Westrum, Results of the first North American comparison of absolute gravimeters, NACAG-2010. J. Geod. 86, 591-596 (2012).
35. B. Malek, Z. Pagel, X. Wu, H. Müller, Embedded control system for mobile atom interferometers. arXiv: 1812.01028 (2018).
36. A. Ryou, J. Simon, Active cancellation of acoustical resonances with an FPGA FIR filter. Rev. Sci. Instrum. 88, 013101 (2017).
37. S. Bade, L. Djadaojee, M. Andia, P. Cladé, S. Cuellati-Khélifa, Photon momentum in a distorted optical field. arXiv:1712.04023v1 (2017).
38. Department of Defense World Geodetic System 1984: Its definition and relationships with local geodetic systems (NIMA Technical Report 8350.2, 3rd ed., National Imagery and Mapping Agency, Washington, DC, USA, 1997).

39. D. Plouff, Preliminary documentation for a FORTRAN program to compute gravity terrain corrections based on topography digitized on a geographic grid (U.S. Geological Survey, Denver, Colorado, 1977).
40. L. L. Nettleton, Determination of density for reduction of gravimeter observations. Geophys. 4, 176-180 (1939).

**Acknowledgement:** We thank Derek van Westrum and Malo Cadoret for carefully reading the manuscript, Dalila Robledo for simulating the magnetic shield, and Anthony Vitan for providing us the design of Campbell Hall and helping to load the apparatus. **Funding:** This work is supported by Bakar Fellows Program at UC Berkeley and NASA Planetary Instrument Definition and Development Program through a Contract with the Jet Propulsion Laboratory. **Author contributions:** X.W., Z.P., B.S.M., F.Z., T.H.N., and H.M. built the apparatus. X.W., Z.P., B.S.M., D.S.S., and H.M. performed the measurements, analyzed the data, and prepared the manuscript. All authors discussed the manuscript. **Competing interests:** The authors declare that they have no competing interests. **Data and materials availability:** All data needed to evaluate the conclusion are present in the paper and the Supplementary Materials. Additional data related to this paper may be requested from the authors.

## Supplementary Materials

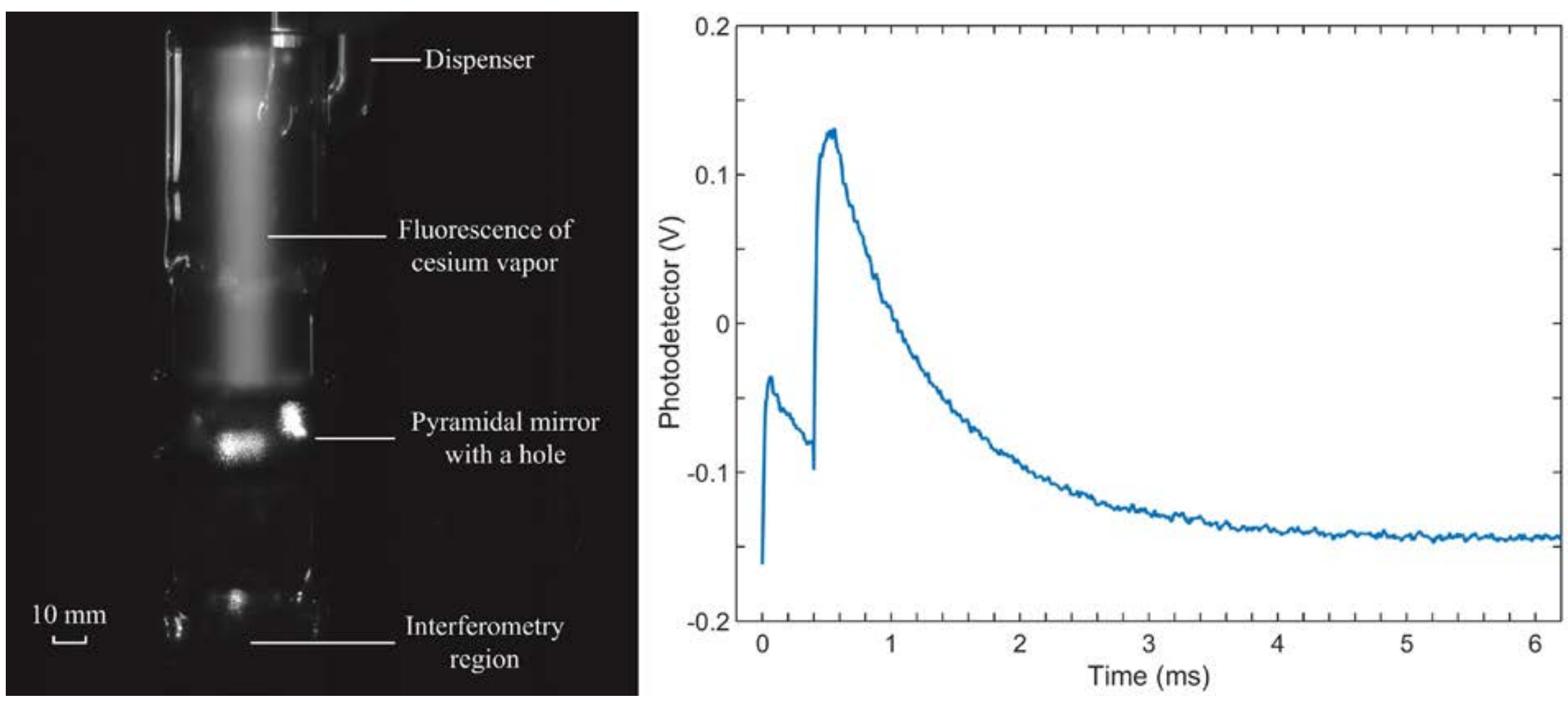

**Fig. S1. Cesium fluorescence inside the glass vacuum chamber.** Left: Vacuum chamber irradiated by the MOT laser beam. The cylinder shape is the glass vacuum chamber. The fluorescence in the MOT region is more than 10-fold stronger than in the interferometry region. Right: Fluorescence detection after atom interferometry. From 0 to 0.4 ms, atoms in $F$=4 state are detected with the cooling beam. From 0.4 to 0.55 ms, the repumping beam is turned on to pump all the atoms to $F$=4 state. From 0.55 to 0.95 ms, all the atoms are detected with the cooling beam. From 5.55 to 5.95 ms, background fluorescence is detected for subtraction.

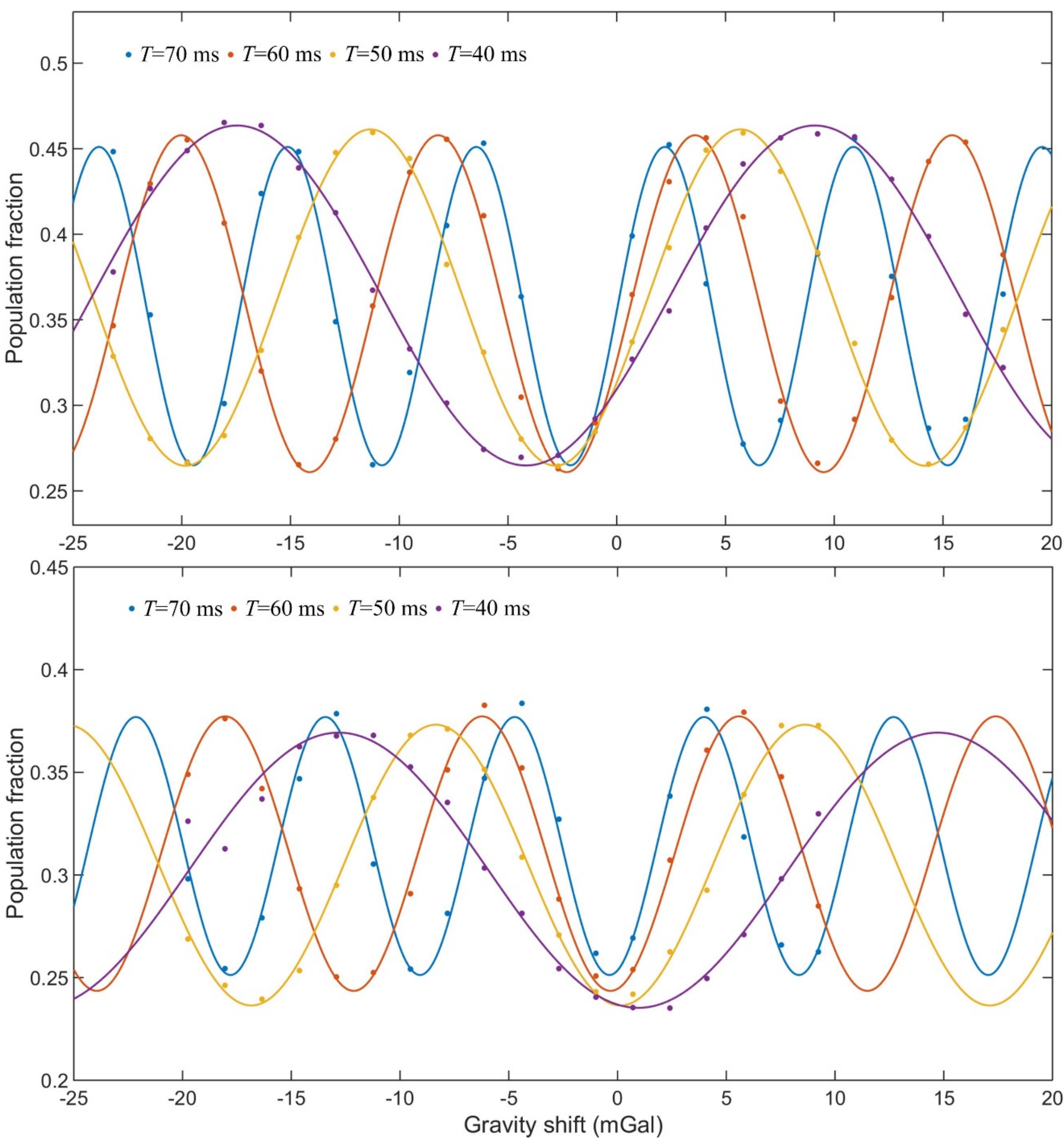

**Fig. S2. Measuring absolute gravity with different pulse separation time.** Top: $k_{eff}$ is positive. Bottom: $k_{eff}$ is negative. The dots are experimental data and the curves are sinusoidal fits. Longer pulse separation time leads to fringes with shorter periods. The overlapped trough is the absolute gravity. The phase shift between fringes with different pulse separation time may be caused by magnetic gradient or AC Stark shift. It can be reduced by reversing the wave vectors.

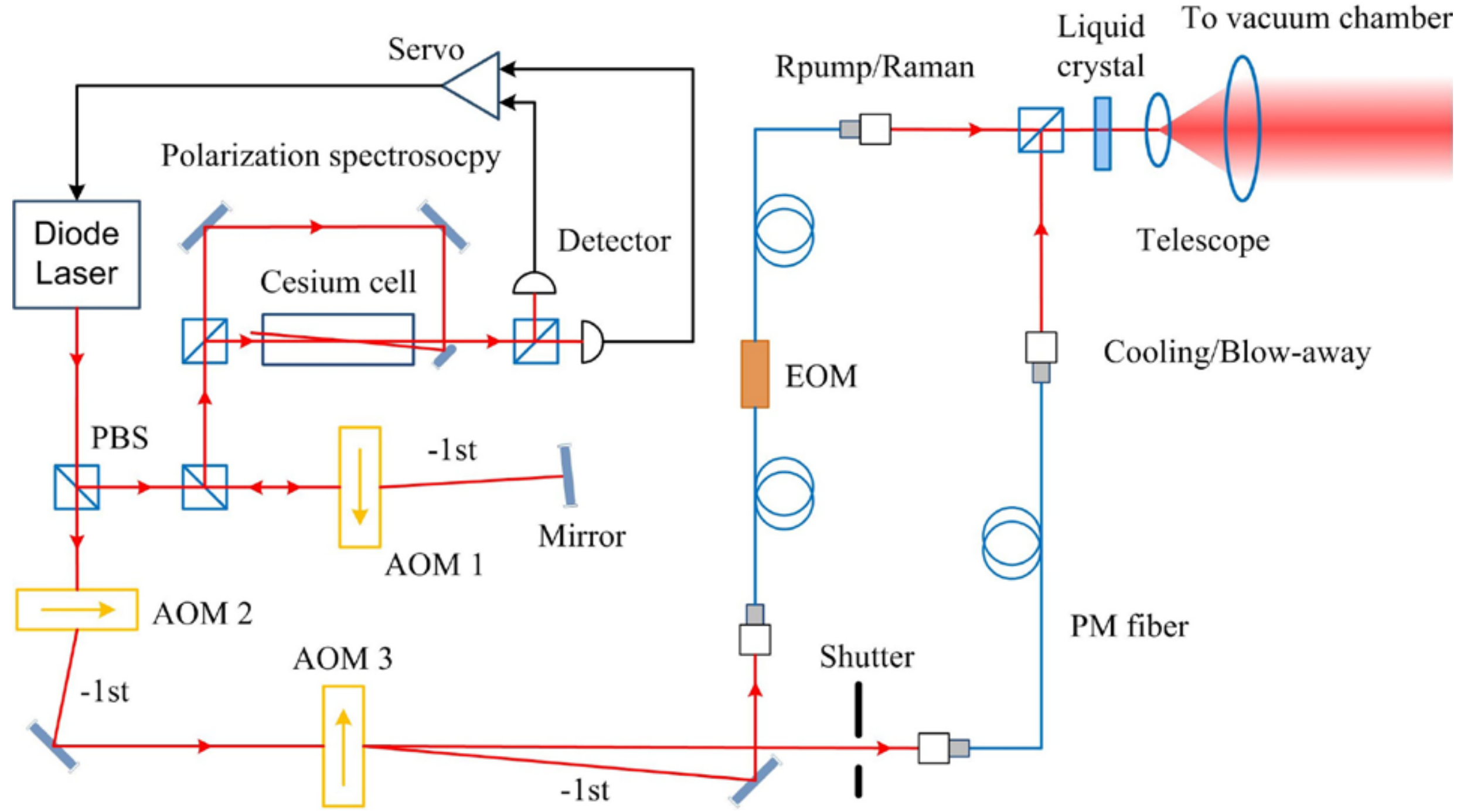

**Fig. S3. Schematic of the laser system.** AOM, acousto-optic modulator; EOM, fiber-based electro-optic modulator; PBS, polarization beam splitter; PM fiber, polarization maintaining fiber.

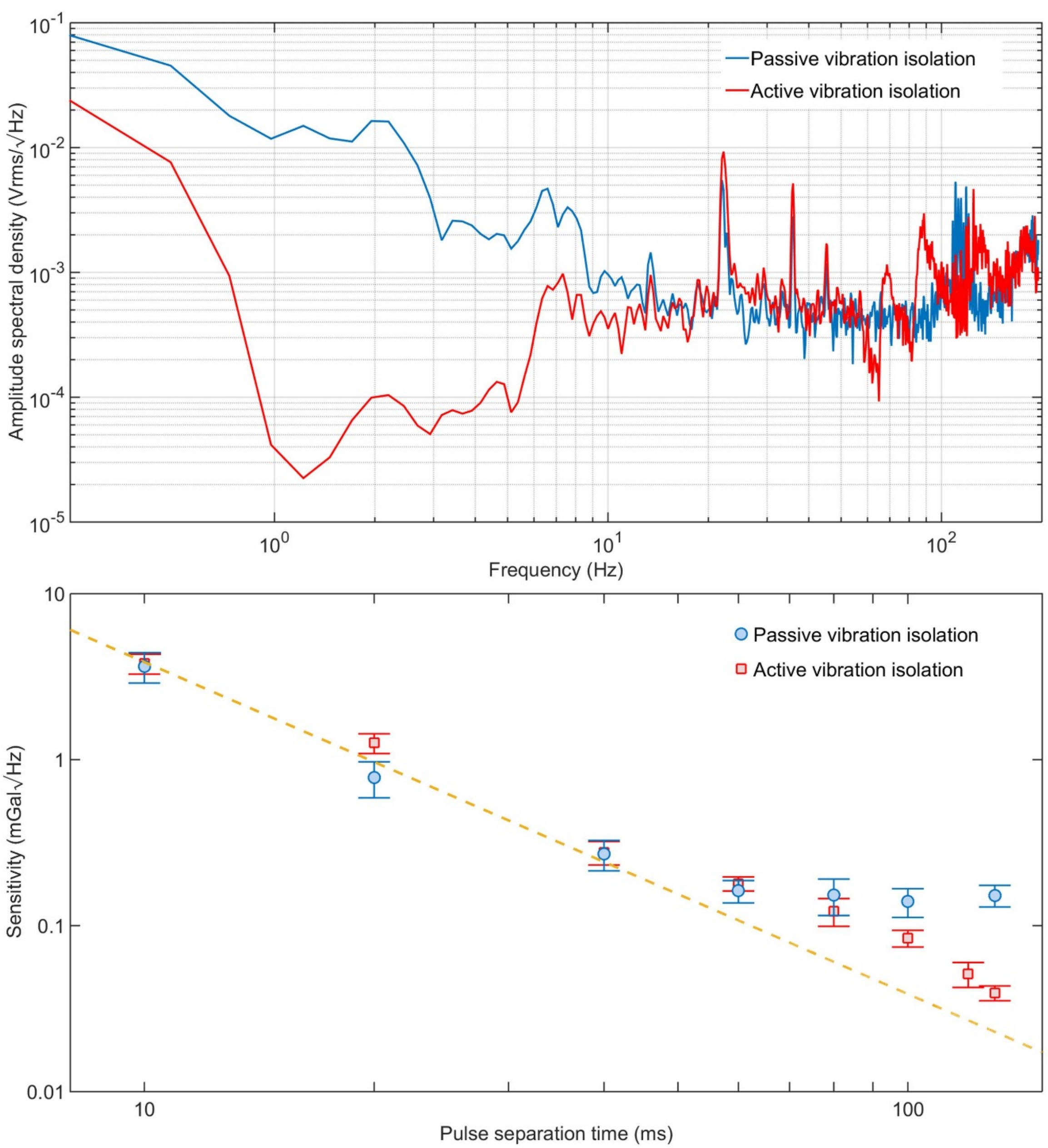

**Fig. S4. Vibration isolation.** Top: Amplitude spectral density. The sensitivity of the seismometer is 750 V/(m/s). The output signal is amplified by a factor of 200 and then measured by a network analyzer. As an in-loop measurement, the active vibration isolation improves the performance by 500-fold around 1 Hz compared with the passive vibration isolation. Bottom: Sensitivity as a function of pulse separation time $T$. The dashed line scales with $T^{-2}$. For $T$=130 ms, the sensitivity using the active vibration isolation is improved by a factor of 5.

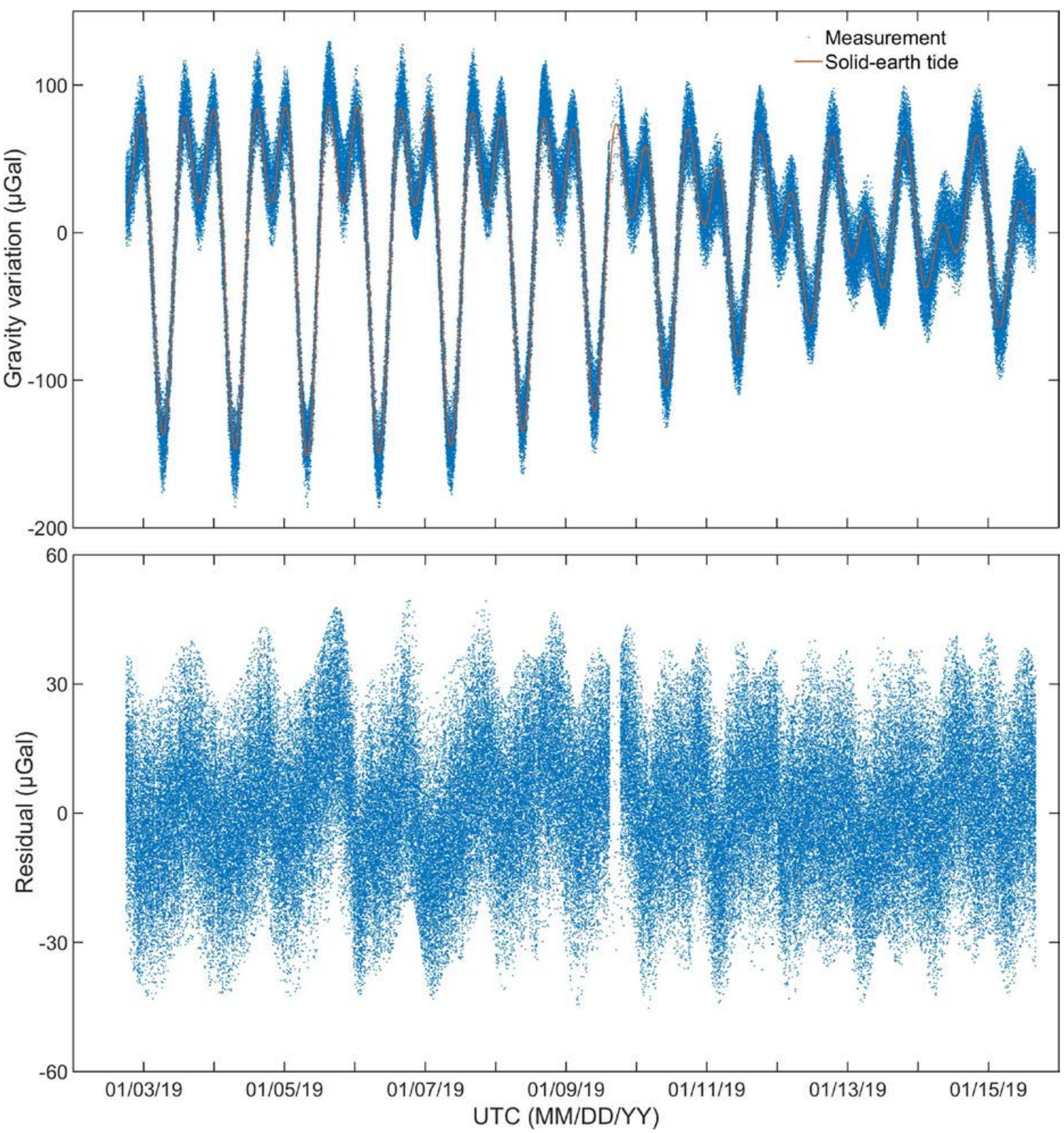

**Fig. S5. Tidal gravity variation.** Top: Comparison between the measured gravity and the solid-earth tide model (30). Bottom: Gravity residual between the measured gravity and the solid-earth tide model. Each dot is one measurement and takes 7.696 s. Data with residuals larger than 50 µGal has been removed, ~0.85% of the total.

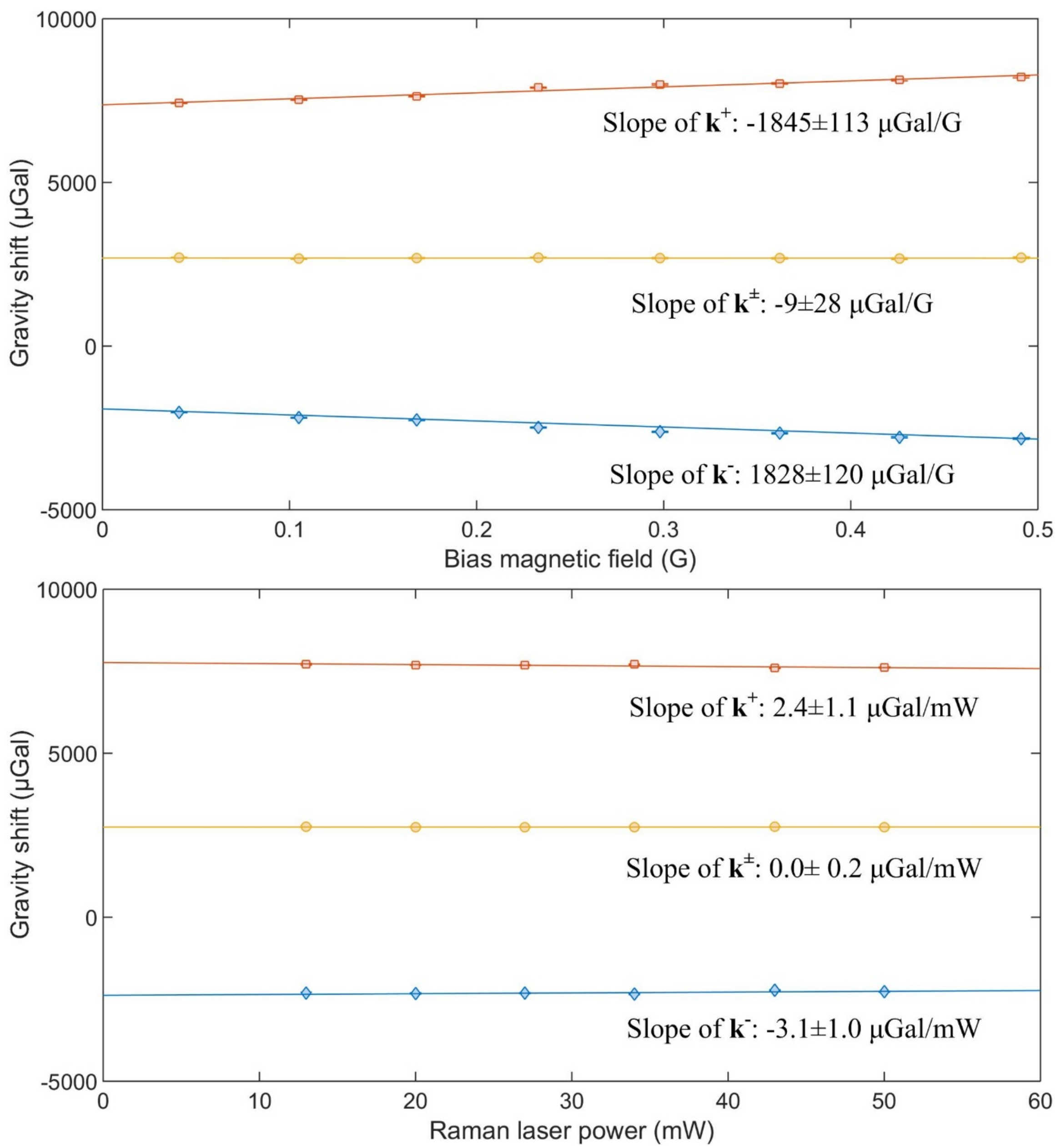

**Fig. S6. Evaluation of systematic effects.** Top: Magnetic fields. Bottom: Differential AC Stark shift. The dots are the experimental data. The error bars are 1-$\sigma$ statistical errors and smaller than the dots. The fit errors are 1-$\sigma$ statistical errors. The lines are linear fits. $\mathbf{k}^+$ is the positive wavevector; $\mathbf{k}^-$ is the negative wavevector; $\mathbf{k}^\pm$ is the average of $\mathbf{k}^+$ and $\mathbf{k}^-$.

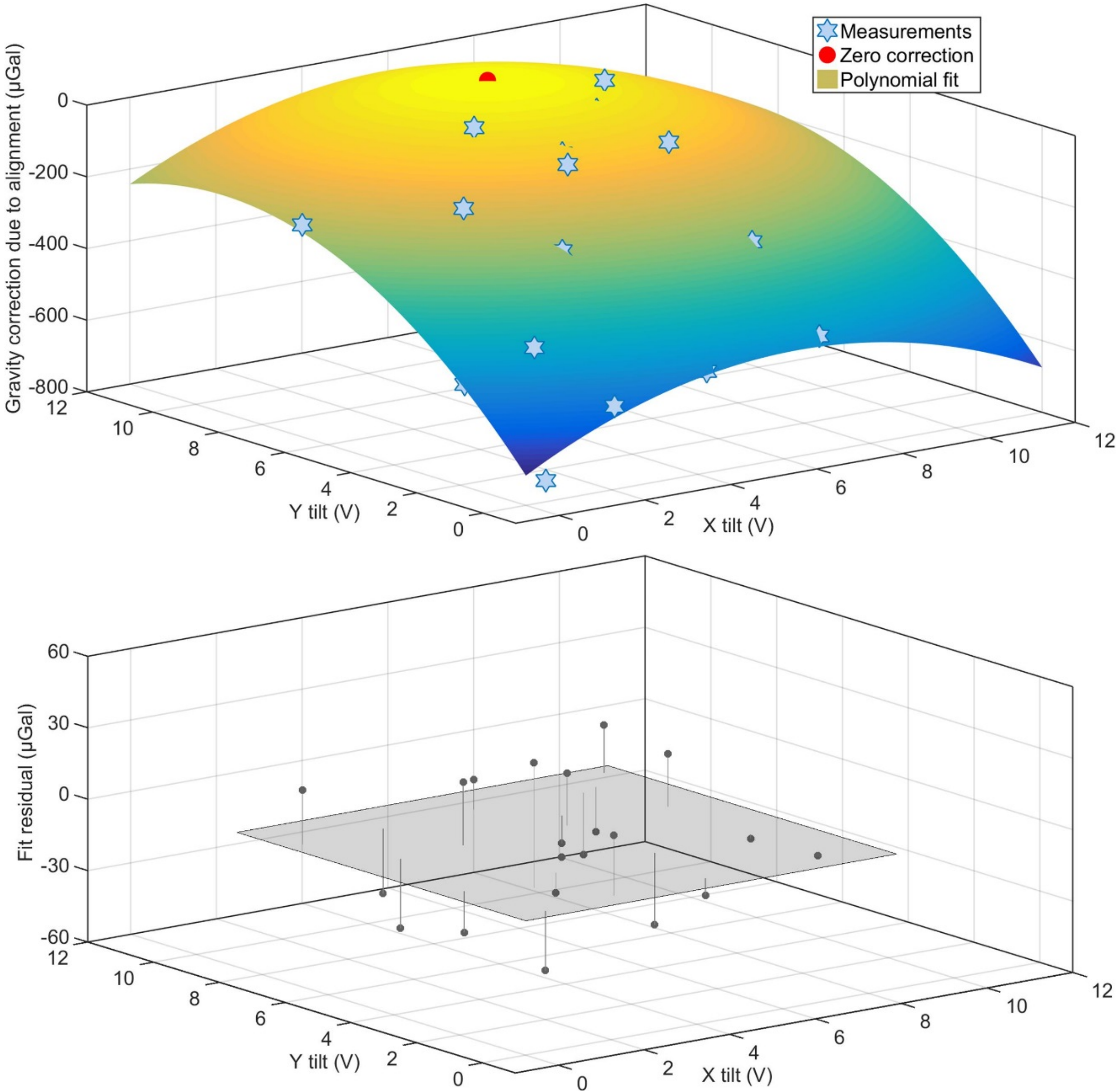

**Fig. S7. Calibration of alignment to gravity axis.** Top: Gravity correction due to tilt of the retroreflector. For the given outputs ($x$, $y$) of the tilt sensor, the bias of the gravity measurements equals ($-717.00+67.34x+113.80y-5.32x^2-0.35xy-6.19y^2$) μGal, where the unit of ($x$, $y$) is volt and the sensitivity of the tilt sensor is 100 μrad/V. Zero correction (red dot) is at the tilt of (6.1, 9.0) V. Bottom: Fit residual. The standard error of the fit residual is ±5 μGal.

**Table S1. Systematic effects.** See Materials and Methods for details.

| Effects | Bias (μGal) | Error (μGal) |
|---|---|---|
| Magnetic fields | -3 | 9 |
| Refractive index of background vapor | -7.3 | 7.3 |
| Coriolis effect | - | 6 |
| Vertical alignment after correction | - | 5 |
| Differential AC Stark shift | - | 3 |
| Raman frequency offset | - | 3 |
| Two-photon light shift | - | 3 |
| Wavefront aberrations | - | 2.1 |
| Gouy phase | - | 1 |
| Laser frequency stability | - | 1 |
| Total | -10.3 | 15 |

**Table S2. Gravity survey in Campbell Hall on UC Berkeley campus.** The height is from the building design. The errors are 1-$\sigma$ statistical and systematic errors. The absolute gravity measured by the atomic gravimeter is corrected to ground-level based on the measured vertical gradient of -0.289 mGal/m and the sensor height of 1.18 m. All gravity values have been corrected for solid-earth tides. The systematic error of the relative gravimeter is ~0.01 mGal due to the uncertainty of the gravity value at the reference station in Menlo Park, California. For each 1-minute measurement with a data update rate of 6 Hz, the standard error of the relative gravimeter is better than 0.005 mGal.

| Level | Height (m) | Atomic gravimeter (mGal) | Relative gravimeter (mGal) |
|---|---|---|---|
| Basement | 0 | 979 955.61(2) | 979 955.58(1) |
| Floor 1 | 4.8768 | 979 954.30(2) | 979 954.25(1) |
| Floor 2 | 9.4488 | 979 952.95(3) | 979 952.95(1) |
| Floor 3 | 13.4112 | 979 951.75(5) | 979 951.84(1) |
| Floor 4 | 17.3736 | 979 950.65(4) | 979 950.71(1) |
| Floor 5 | 21.336 | 979 949.55(3) | 979 949.56(1) |

**Table S3. Gravity survey in Berkeley Hills.** Elevations are from Google maps. The errors are 1-$\sigma$ statistical and systematic errors. The gravity is corrected to ground-level based on the free-air gradient of -0.3086 mGal/m and the sensor height of 2.02 m. Gravity values have been corrected for solid-earth tides, and gravity anomaly values have been further corrected for latitude and terrain effects (see Materials and Methods).

| Longitude (°) | Latitude (°) | Elevation (m) | Gravity (mGal) | Gravity anomaly (mGal) |
|---|---|---|---|---|
| -122.25733 | 37.87277 | 104 | 979 954.81(4) | -26.05(4) |
| -122.25257 | 37.87266 | 123 | 979 949.62(4) | -30.58(4) |
| -122.24691 | 37.87181 | 154 | 979 941.62(2) | -38.24(2) |
| -122.23975 | 37.87563 | 243 | 979 920.71(3) | -59.46(3) |
| -122.24623 | 37.88055 | 353 | 979 897.66(3) | -82.28(3) |
| -122.22231 | 37.88111 | 505 | 979 862.24(6) | -116.52(6) |